\documentclass[journal=jpclett,manuscript=article]{achemso}
\pdfoutput=1
\usepackage{graphicx}
\usepackage{dcolumn}
\usepackage{bm}
\usepackage[colorlinks = true, allcolors = blue]{hyperref}
\usepackage{verbatim}
\usepackage{soul}
\usepackage{float}
\usepackage{xcolor}
\usepackage{chemformula} 
\usepackage[T1]{fontenc} 
\usepackage{multirow}

\author{Diem Thi-Xuan Dang}
\author{Ranjan Kumar Barik}
\author{Manh-Huong Phan}
\author{Lilia M. Woods}
\email{lmwoods@usf.edu.}
\affiliation{Department of Physics, University of South Florida, Tampa, Florida 33620, USA}

\title{Enhanced Magnetism in Heterostructures with Transition Metal Dichalcogenide Monolayers }

\begin{document}

\begin{abstract}
Two-dimensional materials and their heterostructures have opened up new possibilities for magnetism at the nanoscale. In this study, we utilize first-principles simulations to investigate the structural, electronic, and magnetic properties of \ch{Fe/WSe2/Pt} systems containing pristine, defective, or doped \ch{WSe2} monolayers. The proximity effects of the ferromagnetic Fe layer are studied by considering defective and Vanadium-doped \ch{WSe2} monolayers. All heterostructures are found ferromagnetic, and the insertion of the transition metal dichalcogenide results in redistribution of spin orientation and an increased density of magnetic atoms due to the magnetized \ch{WSe2}.  There is an increase in the overall total density of states at the Fermi level due to \ch{WSe2}, however, the transition metal dichalcogenide may lose its distinct semiconducting properties due to the stronger than van der Waals coupling. Spin resolved electronic structure properties are linked to larger spin Seebeck coefficients found in heterostructures with \ch{WSe2} monolayers. 

\end{abstract}

Heterostructures (HSTs) composed of ferromagnetic (FM) and heavy metal (HM) films are typical systems employing spin-charge conversion mechanisms for spintronics applications \cite{Dieny2020}. The quality of such devices is controlled by the atomic composition of the interface and its spin-dependent properties. In this context, HSTs with 2D layered materials have become of great interest \cite{Sierra2021}. Chemically inert layered materials can improve the interface atomic sharpness and they provide new ways of property tuning. For example, semiconducting transition metal dichalcogenides (TMDCs) have strong spin orbit coupling (SOC) which allows the manipulation of spins via electrical means \cite{Xiao2012}. On the other hand, their strong spin photon coupling enables the control of magnetism with optical methods \cite{Phan2021}. The recent discovery of FM ordering at room temperature in  \ch{WS2} and \ch{WSe2} monolayers doped by V atoms \cite{Yun2020, Pham2020, Zhang2020, Jimenez2021} has shown that TMDCs can be effective dilute 2D ferromagnets with robust magnetic states, which can also be light-tunable. 

HSTs with semiconducting TMDC layers offer possibilities for new ways to change and control magnetic properties,  spin transport and spin conversion phenomena through novel interfacial coupling mechanisms. For example, much enhanced magnetic exchange fields, valley splitting, and polarization effects tunable via optical methods have been observed in \ch{WSe2}/\ch{CrI3} HSTs \cite{Zhong2017, Seyler2018, Zhong2020}. Several impressive experimental studies have shown that spin current generation via thermal gradients across FM/TMDC/HM structures is significantly altered when compared with FM/HM systems. Specifically, a large enhancement of the voltage difference due to the longitudinal spin Seebeck effect in \ch{Pt/Ni81Fe19} and Pt/YIG after inserting \ch{WS2} (\ch{WSe2}) layers has been reported \cite{Dastgeer2019, Lee2020, Lee2020_a, Lee2021}. It was also found that such an increase is dependent on the number of layers in the heterostructure, degree of coverage by the TMDC layers, and quality of the interface. Others have examined spin-dependent transport in \ch{Fe3O4/MoS2/Fe3O4} and \ch{Co/Al2O3/MoS2} junctions \cite{Dankert2017, Galbiati2018}. Spin valves based on spin filtering in \ch{WS2/Co} and \ch{NiFe/MoS2/NiFe} have also been reported \cite{Wang2015, Zatko2019}. A gate-tunable magnetic order in \ch{WSe2} by changing the occupation of the Se vacancy states are successfully demonstrated \cite{Nguyen2021} and computational work investigates the origin of ferromagnetism in complex configurations of Se vacancies in V-doped \ch{WSe2} monolayers \cite{Yun2022}. The generated Se vacancies play a key role in the structural change of \ch{PdSe2} because of the laser-induced phase transformation of layered \ch{PdSe2} to metallic phase \ch{PdSe_{2–x}} \cite{Shautsova2019}. Ozyilmaz {\it et. al.} suggested that the spin Hall effect in graphene/TMDC heterostructures may also originate from chalcogenide vacancies \cite{Avsar2014}.

In addition to these experimental demonstrations, the theoretical understanding of the interface properties and coupling mechanisms of FM/TMDC/HM systems must also be advanced. For example, the changes in the band structure  of individual FM, TMDC, and HM layers calculated from first principles can help one understand the overall behavior of the entire HST. Computed charge transfer, bonding, and interfacial hybridization are also necessary in order to highlight magnetic proximity effects unique to FM/TMDC/HM HSTs. Additionally, first principles simulations are necessary to reveal spin selection transfer mechanisms across doped and defective TMDCs. 

In this paper, we present first principles simulations of \ch{Fe/WSe2/Pt} HSTs to elucidate their basic electronic and magnetic properties for the first time. We consider several cases in which the TMDC is a pristine \ch{WSe2}, a monolayer with V doping and a monolayer with various types of vacancies. The interplay between the proximity magnetism from the Fe substrate, the charge transfer and orbital hybdrization across the \ch{WSe2} monolayer and the large SOC effects from the Pt layer leads to a complex picture of interface magnetism. The calculated band structure, as well as the Fermi surface for each component of the considered HSTs, are instrumental for the detailed microscopic analysis of the electronic and spin-dependent behavior across the interface.


\begin{figure}[ht]
    \begin{center}
    \includegraphics[width = 0.9 \textwidth]{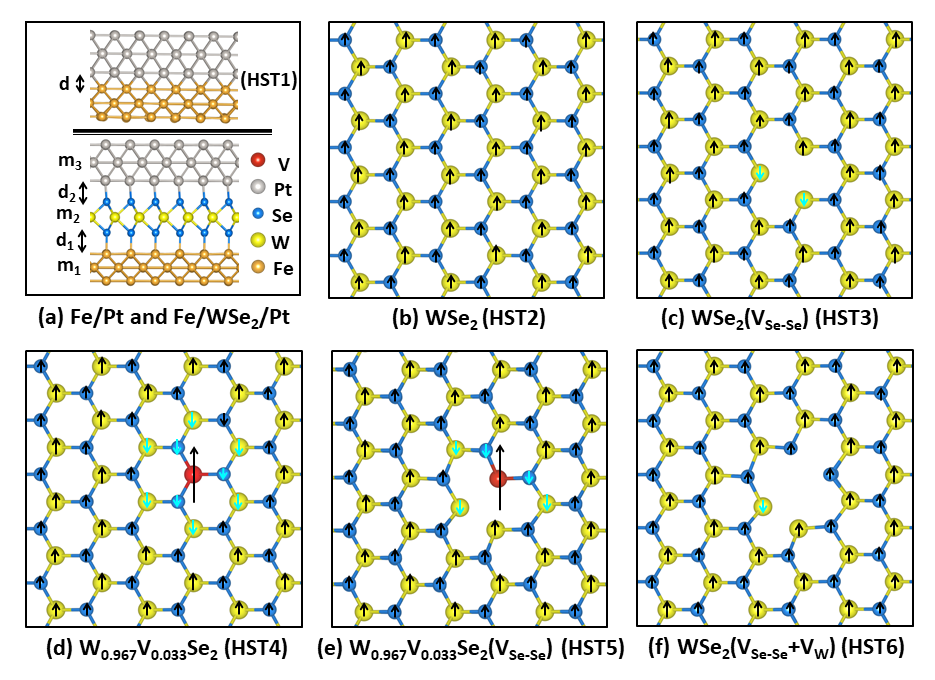}
    \caption{(a) Side view of \ch{Fe/Pt} and \ch{Fe/WSe2/Pt} HSTs. Top views of the various \ch{WSe2} monolayers as part of the \ch{Fe/WSe2/Pt} HSTs: (b) pristine \ch{WSe2} (HST2) (c) \ch{WSe2}(\ch{V_{Se-Se}}), a defective monolayer with a pair of Se vacancies (HST3); (d) \ch{W_{0.967}V_{0.033}Se2}, a defective monolayer with 3.3\% V doping substituting W (HST4); (e) \ch{W_{0.967}V_{0.033}Se2}(\ch{V_{Se-Se}}), a defective monolayer with 3.3\% V doping and a pair of Se vacancies (HST5); (f) \ch{WSe2}(\ch{V_{Se-Se}} + \ch{V_W}), a defective monolayer with a pair of Se vacancies and a W vacancy (HST6). The interlayer distances are denoted in panel (a), while the electronic spin directions for all atoms of the monolayers are also shown with arrows upon completing the spin-polarized calculations.}
    \label{fig:1}
    \end{center}
\end{figure}

Different HSTs have been constructed by taking the middle layer as pristine \ch{WSe2}, as well as monolayers with V atom substitutional dopants or Se or W vacancies. A side view of the Fe/Pt HST (taken as a reference) and a HST with a \ch{WSe2} monolayer inserted in-between the Fe and Pt layers is shown in Fig. \ref{fig:1}a.  As part of the \ch{Fe/WSe2/Pt} HST, we consider a perfect \ch{WSe2} monolayer, a monolayer with two Se vacancies - \ch{WSe2} (\ch{V_{Se-Se}}), a monolayer with 3.3\% doping of V atoms substituting W atoms – \ch{W_{0.957}V_{0.033}Se2}, a  monolayer with two Se vacancies and a 3.3\% V doping - \ch{W_{0.957}V_{0.033}Se2} (\ch{V_{Se-Se}}), and a monolayer containing two Se and one W vacancies – \ch{WSe2} (\ch{V_{Se-Se}} + \ch{V_W}), as shown in Fig. \ref{fig:1}b-f. 

In each case, the TMDC monolayer is in a 2H configuration. Fe and Pt layers with $(001)$ surface termination, have been simulated to create the HSTs having layer thickness 2.804 \AA \text{ } and 3.926 \AA, respectively. For the calculations, a supercell consisting of $6 \times 6$ Fe unit cells (108 atoms), $3\sqrt{3} \times 5$ unit cells for the perfect monolayer (90 atoms), and $3\sqrt{2} \times 3\sqrt{2}$ Pt unit cells (108 atoms) is constructed. In the case of the reference \ch{Fe/Pt} HST, the TMDC monolayer is removed, and for the others, defective and/or doped \ch{WSe2} are part of each HST (as shown in Fig. \ref{fig:1}). 

\begin{table}[ht]
\centering
\resizebox{\textwidth}{!}{
\caption{Basic properties of the considered HSTs from Fig. \ref{fig:1}. The strain $\epsilon$ (\%) along the inequivalent $a, b$ lattice directions with and without the vdW correction is shown for the layer with the largest shrinkage.  The interlayer separations $d$, $d_1$, $d_2$ are also given with and without vdW interaction. The binding energy is calculated using the expression $E_b=\left(E_{HST}-\sum\limits_{i} E_i^{layer} \right)/N$, where $E_{HTS}$ is the total energy of the HST, $E_i^{layer}$ is the total energy of the $i^th$  isolated layer, and $N$ is the number of atoms in the HST unit cell. The Bader charge is shown for each layered component of the HSTs. The magnetization per atom $m_1$, $m_2$, $m_3$ corresponds to each layer and $M$ is the total average magnetization of each HST. The Bader charges and magnetizations are obtained with the vdW correction taken into the calculations.}
\label{tab:1}
\begin{tabular}{|c|rr|cc|cc|ccc|r|c|c|c|}
\hline\hline
\multirow{2}{*}{} & \multicolumn{2}{c|}{Strain (\%)}                                                                                                       & \multicolumn{2}{c|}{Layer seperation (\AA)}                                                                                                     & \multicolumn{2}{c|}{$E_b$ (meV)}         & \multicolumn{3}{c|}{Bader charge $(e)$}                                                  & \multicolumn{1}{c|}{\multirow{2}{*}{\begin{tabular}[c]{@{}c@{}}$m_1$\\ ($\mu_B$)\end{tabular}}} & \multicolumn{1}{c|}{\multirow{2}{*}{\begin{tabular}[c]{@{}c@{}}$m_2$\\ ($\mu_B$)\end{tabular}}} & \multicolumn{1}{c|}{\multirow{2}{*}{\begin{tabular}[c]{@{}c@{}}$m_3$\\ ($\mu_B$)\end{tabular}}} & \multicolumn{1}{c|}{\multirow{2}{*}{\begin{tabular}[c]{@{}c@{}}$M$\\ ($\mu_B$)\end{tabular}}}
\\ \cline{2-10}
                  & \multicolumn{1}{c|}{No vdW}                                                  & \multicolumn{1}{c|}{vdW}                                & \multicolumn{1}{c|}{No vdW}                                                              & vdW                                                                 & \multicolumn{1}{c|}{No vdW}   & vdW      & \multicolumn{1}{c|}{Fe}    & \multicolumn{1}{c|}{TMDC}   & \multicolumn{1}{c|}{Pt} & \multicolumn{1}{c|}{}                                                                           & \multicolumn{1}{c|}{}                                                                           & \multicolumn{1}{c|}{} 
                                                                           & \multicolumn{1}{c|}{} \\ \hline\hline
HST1              & \multicolumn{1}{r|}{\begin{tabular}[c]{@{}r@{}} (a, Fe) -3.428\\ (b, Fe) -3.411\end{tabular}} & \begin{tabular}[c]{@{}r@{}}-3.629\\ -3.620\end{tabular} & \multicolumn{1}{c|}{$d= 1.795$}                                                          & $d= 1.787$                                                          & \multicolumn{1}{c|}{-416.247} & -509.738 & \multicolumn{1}{r|}{4.163} & \multicolumn{1}{c|}{--}     & -4.163                  & 2.579                                                                                           & \multicolumn{1}{c|}{--}                                                                         & -0.058 & 2.521                                                                                          \\ \hline
HST2              & \multicolumn{1}{r|}{\begin{tabular}[c]{@{}r@{}}(a, Fe) -2.045\\ (b, Fe) -3.460\end{tabular}} & \begin{tabular}[c]{@{}r@{}}-2.563\\ -3.800\end{tabular} & \multicolumn{1}{c|}{\begin{tabular}[c]{@{}c@{}}$d_1= 2.405$\\ $d_2= 2.405$\end{tabular}} & \begin{tabular}[c]{@{}c@{}}$d_1= 2.284$\\ $d_2= 2.289$\end{tabular} & \multicolumn{1}{c|}{-86.078}  & -195.852 & \multicolumn{1}{r|}{0.888} & \multicolumn{1}{r|}{0.888}  & -1.776                  & 2.588                                                                                           & 0.014                                                                                           & 0.023  & 2.625
                                                                                \\ \hline
HST3              & \multicolumn{1}{r|}{\begin{tabular}[c]{@{}r@{}}(a, Fe) -2.300\\ (b, Fe) -3.601\end{tabular}} & \begin{tabular}[c]{@{}r@{}}-2.774\\ -3.906\end{tabular} & \multicolumn{1}{c|}{\begin{tabular}[c]{@{}c@{}}$d_1= 2.363$\\ $d_2= 2.379$\end{tabular}} & \begin{tabular}[c]{@{}c@{}}$d_1= 2.200$\\ $d_2= 2.279$\end{tabular} & \multicolumn{1}{c|}{-87.749}  & -197.653 & \multicolumn{1}{r|}{2.368} & \multicolumn{1}{r|}{-1.143} & -1.226                  & 2.589                                                                                           & 0.012                                                                                           & 0.018  &2.619                                                                                         \\ \hline
HST4              & \multicolumn{1}{r|}{\begin{tabular}[c]{@{}r@{}}(a. Fe) -1.994\\ (b, Fe) -3.392\end{tabular}} & \begin{tabular}[c]{@{}r@{}}-2.510\\ -3.731\end{tabular} & \multicolumn{1}{c|}{\begin{tabular}[c]{@{}c@{}}$d_1= 2.244$\\ $d_2= 2.320$\end{tabular}} & \begin{tabular}[c]{@{}c@{}}$d_1= 2.143$\\ $d_2= 2.234$\end{tabular} & \multicolumn{1}{c|}{-88.237}  & -199.307 & \multicolumn{1}{r|}{3.450} & \multicolumn{1}{r|}{-2.796} & -1.822                  & 2.589                                                                                           & 0.026                                                                                           & 0.022    &2.637                                                                                       \\ \hline
HST5              & \multicolumn{1}{r|}{\begin{tabular}[c]{@{}r@{}}(a, Fe) -2.540\\ (b, Fe) -3.432\end{tabular}} & \begin{tabular}[c]{@{}r@{}}-2.717\\ -3.777\end{tabular} & \multicolumn{1}{c|}{\begin{tabular}[c]{@{}c@{}}$d_1= 2.311$\\ $d_2= 2.282$\end{tabular}} & \begin{tabular}[c]{@{}c@{}}$d_1= 2.064$\\ $d_2= 2.209$\end{tabular} & \multicolumn{1}{c|}{-91.638}  & -201.194 & \multicolumn{1}{r|}{3.562} & \multicolumn{1}{r|}{-1.004} & -2.559                  & 2.594                                                                                           & 0.040                                                                                           & 0.025   &2.659                                                                                        \\ \hline
HST6              & \multicolumn{1}{r|}{\begin{tabular}[c]{@{}r@{}}(a, Fe) -2.255\\ (b, Fe) -3.358\end{tabular}} & \begin{tabular}[c]{@{}r@{}}-2.758\\ -3.731\end{tabular} & \multicolumn{1}{c|}{\begin{tabular}[c]{@{}c@{}}$d_1= 2.180$\\ $d_2= 2.191$\end{tabular}} & \begin{tabular}[c]{@{}c@{}}$d_1= 2.043$\\ $d_2= 2.120$\end{tabular} & \multicolumn{1}{c|}{-98.094}  & -207.405 & \multicolumn{1}{r|}{4.196} & \multicolumn{1}{r|}{-1.001} & -3.195                  & 2.596                                                                                           & 0.016                                                                                           & 0.025  &2.637                                                                                        \\ \hline\hline
\end{tabular}
}
\end{table}

The formation of the supercells introduces slight strains when compared with the lattice structures of the individual free-standing constituents. The strain $\epsilon$ is quantified using $\epsilon = \frac{a_{HST} - a_i}{a_i}  \times 100 \% \; $  ($a_{HST}$ - lattice constant of the HST; $a_i$ - lattice constant of the Fe, TMDC, or Pt layers). In Table \ref{tab:1}, we show results for $\epsilon$ of the Fe layer along the inequivalent $a$ and $b$ lattice directions, and all data is given in Table S-1 in the Supporting Information. For the Fe and Pt layers, all values are negative indicating that the lattice has shrunk, and overall this  effect is more pronounced when the vdW interaction is taken into account. On the other hand, slight stretching along the $b$ direction and shrinkage along the $a$ direction are found for all considered $\ch{WSe2}$ monolayers. Nevertheless, $|\epsilon|$ does not exceed $ \sim 3.9\%$.  

Interlayer separations for each HST  with and without the vdW corrections are also reported in Table \ref{tab:1}. Including the vdW interaction in the simulations results in decreased $d_1$, $d_2$ as expected. In fact, the interlayer distances are smaller than the typical vdW separations (usually $\succeq 3$ \text{ \AA}), which is indicative of stronger interactions between the layers. Our results for the calculated binding energies (with included vdW correction), also given in Table \ref{tab:1}, show that $E_b \sim -500$ meV for the reference Fe/Pt HST is increased to $E_b \sim -200$ meV for HSTs with a \ch{WSe2} monolayer, which is expected since the chemically inert  TMDC monolayer increases the separation between Fe and Pt. Nevertheless, $|E_b| \sim 200$ meV suggests that stronger chemical effects coexist with the vdW interaction typically associated with interlayer coupling of TMDCs \cite{Pollard2016}. 

To elucidate the chemical interaction in the HSTs, the Bader charges (expressed as atomic charges in relation to the neutral ones) for each layer are also calculated and given in Table \ref{tab:1}. The Bader charge is a reliable quantitative measure for the charge distribution in a given system \cite{Henkelman2006, Koch2018}. From the numerical data in Table \ref{tab:1}, we find that the Fe layer and the perfect \ch{WSe2} take equal amounts of positive charge from Pt. However, this balance is changed in the case of defective or doped \ch{WSe2} whose charge becomes negative, while the Fe layer acquires an enhanced positive charge. In all cases the Pt layer charge is always negative, but its magnitude is reduced when comparing with the Fe/Pt HST and the systems containing the TMDC layer. This is well correlated with the shorter $d$ and stronger $E_b$, as compared to $d_1$, $d_2$ and the binding energy for the HSTs involving a TMDC layer.

The electron gain or loss between the different components in the HSTs is intrinsically related to the atomic magnetic moment distribution, also given in Fig. \ref{fig:1} and Table \ref{tab:1}. The average magnetic moments $m_1, m_2, m_3$ after relaxation show how the spin polarization for each layer changes as different HSTs are constructed. In all cases, the calculated magnetic moments are along the direction perpendicular to the layer surfaces featuring FM polarization. For the reference Fe/Pt system, the overall magnetization is always positive, however, the Pt layer itself has opposite spin polarization compared to the one for Fe. In the \ch{Fe/WSe2/Pt} (HST2), each atom from the TMDC monolayer is spin-polarized in the upward direction with an overall layer magnetization $0.014$ $\mu_B$ due to the proximity of the Fe layer. It is interesting to note that $m_1$ for the Fe layer has increased, while the Pt layer has reversed its spin polarization direction with a reduced by a factor of two magnitude when compared to the Fe/Pt HST. Nevertheless, the total magnetization in HSTs with \ch{WSe2} monolayers is enhanced when compared to the reference \ch{Fe/Pt}. The effect is most pronounced for the case with the V doped defective monolayer due to all $m_1, m_2,$ and $m_3$ being the largest, as seen in  Table \ref{tab:1}.

The spin direction of each atom (depicted in Fig. \ref{fig:1})   in the doped TMDC layer further shows that the V atom and its surrounding first and second nearest neighbors, even in the presence of Se vacancies, are coupled antiferromagnetically. This type of rather local AFM frustration is quenched, however, showing a preferred FM orientation over a longer period \cite{Zhang2020}. On the other hand, all neighboring atoms in the presence of W and Se vacancies are FM coupled to their neighbors, as evident in HST6. The results indicate that the defective or doped HSTs have similar $m_1$, $m_2$, $m_3$, with the largest $m_2$ is observed for the \ch{W_{0.967}V_{0.033}Se2(V_{Se-Se})} from HST5 (Table \ref{tab:1}).

\begin{figure}[H]
    \begin{center}
    \includegraphics[width = 1 \textwidth]{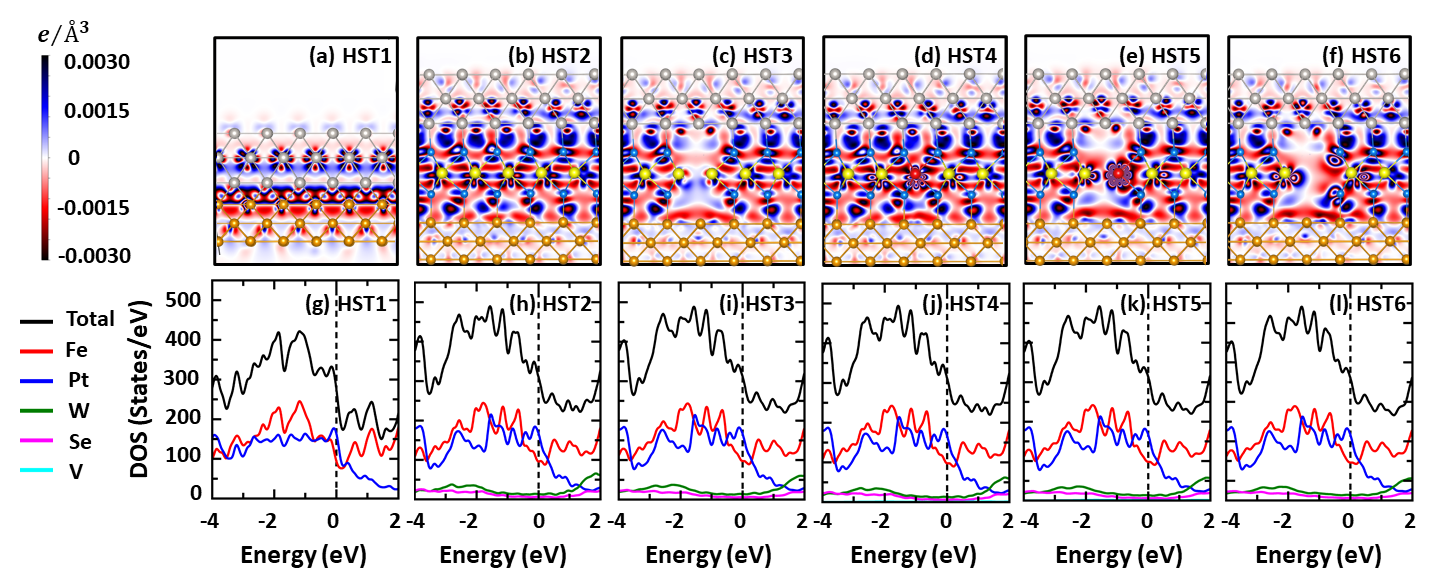}
    \caption{Charge transfer in (a) Fe/Pt (HST1), (b) \ch{Fe/WSe2/Pt} (HST2), (c) \ch{Fe/WSe2 (V_{Se-Se})/Pt} (HST3), (d) \ch{Fe/W_{0.967}V_{0.033}Se2/Pt} (HST4), (e) \ch{Fe/W_{0.967}V_{0.033}Se2 (V_{Se-Se})/Pt} (HST5), (f) \ch{Fe/WSe2 (V_{Se-Se} + V_W)/Pt} (HST6). The color coding for the atoms is the same as in Fig. \ref{fig:1}. Total and atomically resolved density of states (DOS) for (g) HST1, (h) HST2, (i) HST3, (j) HST4, (k) HST5 and (l) HST6.}
    \label{fig:2}
    \end{center}
\end{figure}

An important factor for the bonding in the HSTs is the charge transfer, which can be calculated using $ \delta \rho_{HST} = \rho_{HST} - ( \rho_{Fe} + \rho_{WSe_2} + \rho_{Pt}) \text{ } $ whereas $\rho_{HST}$, $\rho_{Fe}$, $ \rho_{WSe_2}$, $\rho_{Pt}$ are the total charge in the HST, isolated Fe layer, isolated \ch{WSe2} monolayer and the isolated Pt layer, respectively. In Fig. \ref{fig:2}, $\delta \rho_{HS}$  is shown for the considered HSTs from Fig. \ref{fig:1}. Charge is moved from depleted regions (red isosurfaces) to regions with accumulated electrons (blue isosurfaces). A uniform layer of electron accumulation corresponds to ionic-like bonding as found at the Fe/Pt boundary indicating strong chemical interaction between the two layers (Fig. \ref{fig:2}a). Upon insertion of the \ch{WSe2} monolayer, localized depletion above the Se atoms and localized electron accumulation above Pt and Fe atoms is observed (Fig. \ref{fig:2}b). These ellipsoidal-like regions suggest dipole-dipole charge density distribution. The presence of Se vacancies (Fig. \ref{fig:2}c) or V dopants (Fig. \ref{fig:2}d) does not significantly affect the Fe/Se and Pt/Se interface as the main changes are primarily at the location of these disturbances. Comparing the different cases displayed in Fig. \ref{fig:2} shows that the charge transfer is rather concentrated at the direct Fe/Pt interface, and the \ch{WSe2} monolayer serves as a medium through which charge is transferred weakens the bond between Fe and Pt layers. 

In Fig. \ref{fig:2}g-l, the calculated density of states (DOS) is also shown. In the case of Fe/Pt, both types of atoms contribute to the DOS around the Fermi level. This situation does not change for the HSTs with \ch{WSe2} monolayers. However, we find that the presence of the TMDC always increases the total DOS at the Fermi level by about $23\%$ regardless of the presence of defects or dopants. This increase in conduction states results in a reduced conductivity mismatch between Fe and Pt, which can have a favorable effect on the enhancement of spin dependent transport phenomena in such systems, as observed experimentally \cite{Dastgeer2019, Lee2020, Lee2020_a, Lee2021}.

Further insight into the microscopic picture of such heterostructures can be gained by considering their energy band structures. Since the different layers are incommensurate, the constructed supercell contains several unit cells from each component. This situation also occurs in the case of monolayer \ch{WSe2} with impurities or V dopants, for which the supercell contains several unit cells. Because of the shrunk Brillouin zone (BZ) appearing in materials containing multiple unit cells, artificially folded bands are created. Thus, we are faced with the problem of band untangling on the primitive BZs for each layer to have a complete analysis of the electronic structure of the various HSTs. As representative examples, we consider Fe/Pt, \ch{Fe/WSe2/Pt} and \ch{Fe/W_{0.9}V_{0.1}Se2/Pt} to compare how the unpolarized perfect \ch{WSe2} and FM V-doped TMDC layer affect the band structure of the reference Fe/Pt.

\begin{figure}[H]
    \begin{center}
    \includegraphics[width = 1 \textwidth]{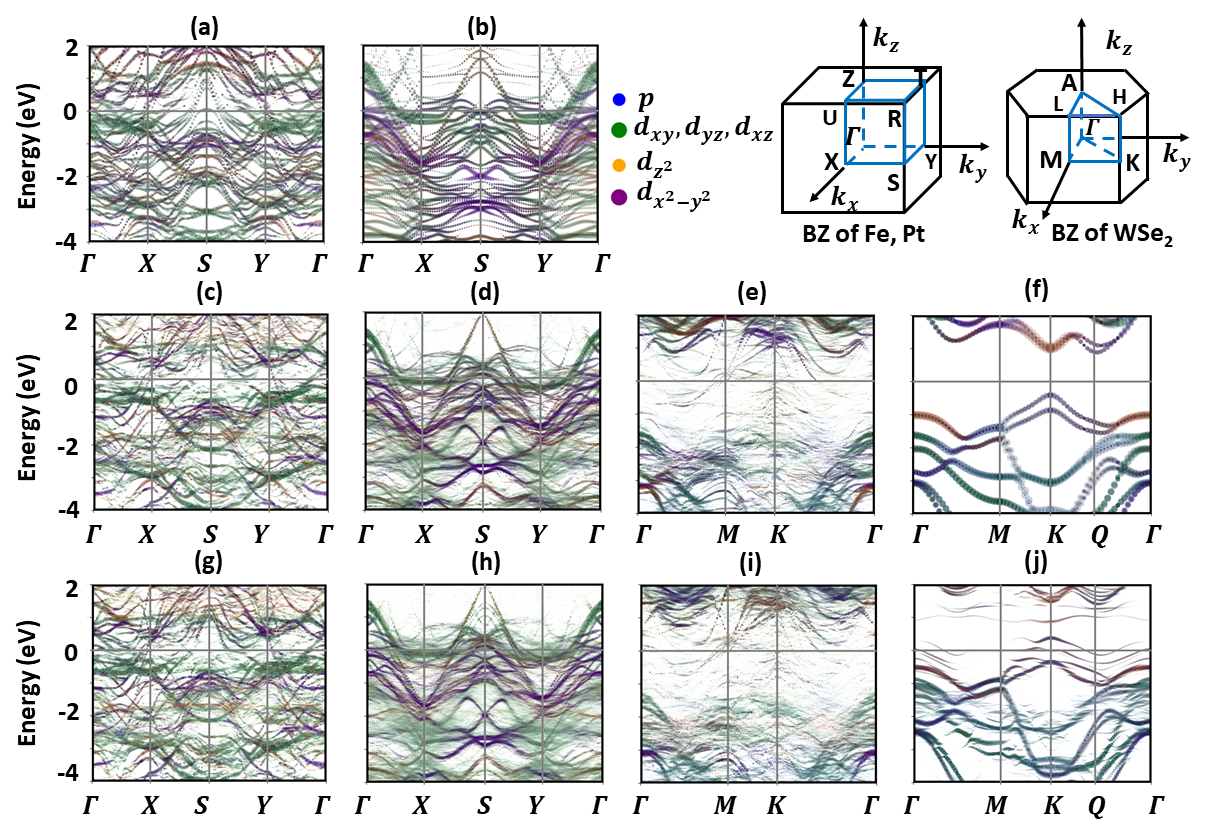}
     \caption{Unfolded band structures for the: (a) Fe layer and (b) Pt layer from the Fe/Pt HST; (c) Fe layer, (d) Pt layer and (e) \ch{WSe2} monolayer from the \ch{Fe/WSe2/Pt} HST; (f) standalone \ch{WSe2} monolayer; (g) Fe layer, (h) Pt layer, and (i) \ch{W_{0.9}V_{0.1}Se2} monolayer from the \ch{Fe/W_{0.9}V_{0.1}Se2/Pt} HST; (j) standalone \ch{W_{0.9}V_{0.1}Se2} monolayer. The band structures are given in the native Brillouin zones for each material. The Brillouin zones (BZ) and high-symmetry points for the Fe, Pt and \ch{WSe2} layers are also shown.}
    \label{fig:3}
    \end{center}
\end{figure}

In Fig. \ref{fig:3}, we show the unfolded bands (obtained as discussed in Methods) as projected on each component native BZ along high symmetry points. For the Fe and Pt layers, the high symmetry points of $\Gamma$ (0 0 0) – X (0.5 0 0) – S (0.5 0.5 0) – Y (0 0 0.5) – $\Gamma$ are selected to calculate the energy bands. Fig. \ref{fig:3}a, b show the orbitally projected bands for both layers. We find that the bands around the Fermi level along the $X - S - Y$ (Fig. \ref{fig:3}a) and $\Gamma - X - S$ (Fig. \ref{fig:3}b) paths are primarily composed of hybridized $d_{xy}$, $d_{yz}$, $d_{xz}$ orbitals. Along $\Gamma - X$ and $Y - \Gamma$ additional hybridization with $d_{z^2}$ and $d_{x^2-y^2}$ is also found. In the Supproting Information, a separate $p$; $d_{xy}$, $d_{yz}$, $d_{xz}$; $d_{z^2}$; $d_{x^2-y^2}$ band decomposition is given for easier viewing together with the unfolded band structures of the separate \ch{Fe/WSe_2} (Fig. S2) and \ch{Pt/WSe_2} (Fig. S3). Upon inserting \ch{WSe2}, the orbital character of the bands around $E_F$ does not change, however, the Fe orbitals become less concentrated, while the contribution from the Pt orbitals and their intensity increase significantly (Fig. \ref{fig:3}c, d, g, h). 

It is further interesting to analyze the \ch{WSe2} monolayer: as a free-standing material, it is an indirect bandgap semiconductor with a gap of $1.390$ eV at the K – Q path, as shown in Fig. \ref{fig:3}f, which is in good agreement with previous studies \cite{Zhang2015, Hu2019}. The emergence of an indirect gap is a consequence of the SOC which is responsible for $d$-band splitting and lowering $d_{x^2-y^2}$ in the conduction region at the Q-point in Fig. \ref{fig:3}f. The band structure of the monolayer changes significantly upon its insertion between Fe and Pt. Fig. \ref{fig:3}e shows several bands crossing $E_F$ and a series of $d$-bands of various intensities around the K-point. It appears that the \ch{WSe2} monolayer has lost its semiconducting properties due to the stronger than vdW bonding with Fe and Pt, as also evidenced from Fig. \ref{fig:3}e. The standalone doped monolayer \ch{W_{0.9}V_{0.1}Se2}, on the other hand, shows an upward shift of the valence bands touching the Fermi level at the K-point. There is further splitting at the top of the valence maximum crossing over to the gap of the monolayer (Fig. \ref{fig:3}j). Impurity levels originating from the V d-states are also found localized near the conduction band minimum. Nevertheless, the band structure of the \ch{W_{0.9}V_{0.1}Se2} as part of the HST in Fig. \ref{fig:3}i washes away these signature features; instead, similar to \ch{WSe2} from \ch{Fe/WSe2/Pt}  there is a series of bands near and at $E_F$ with various intensities  (Fig. \ref{fig:3}e). The Fe and Pt layers, however, result in a similar situation as for the \ch{Fe/WSe2/Pt} HST with a series of bands near and at $E_F$ with various intensities. 

\begin{figure}[H]
    \begin{center}
    \includegraphics[width = 0.9 \textwidth]{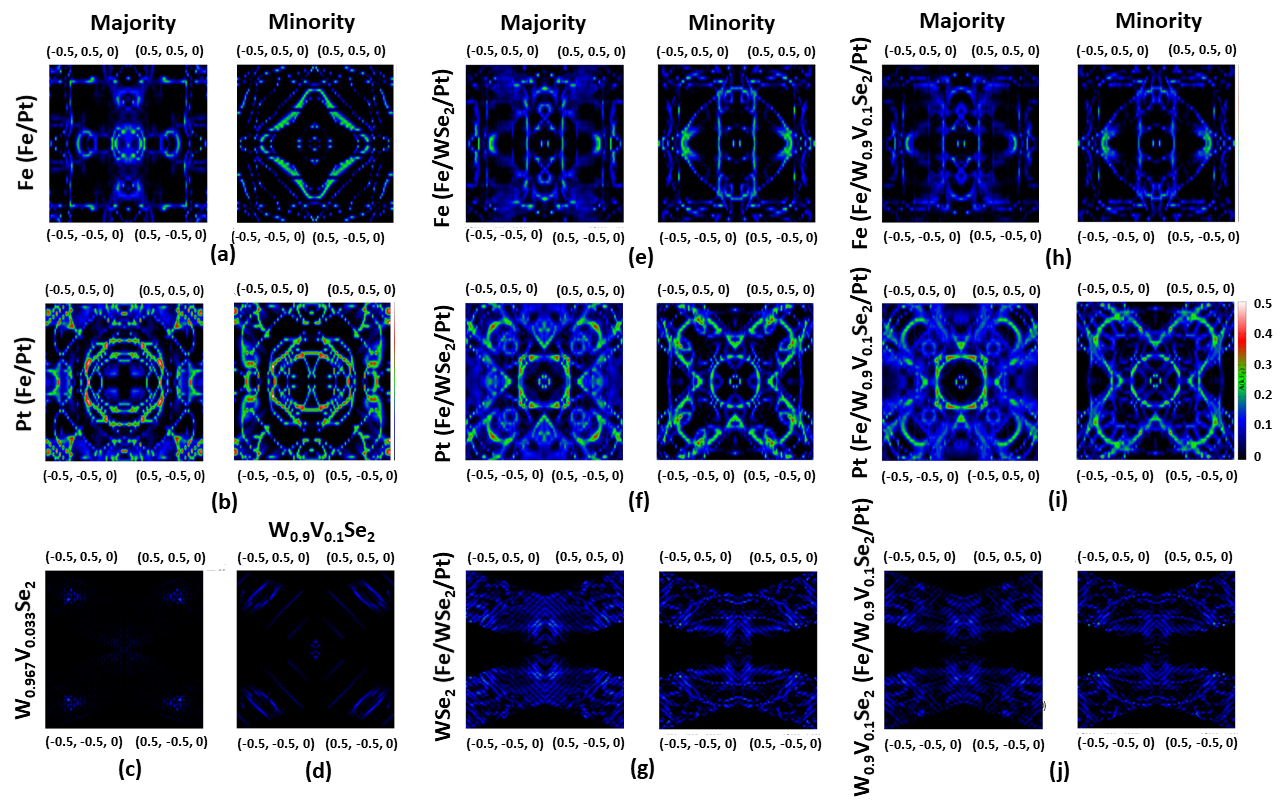}
    \caption{The unfolded Fermi Surface for the: (a) Fe layer and (b) Pt layer from the Fe/Pt heterostructure; (c) \ch{W_{0.967}V_{0.033}Se2} monolayer; (d) \ch{W_{0.9}V_{0.1}Se2} monolayer; (e) Fe layer, (f) Pt layer and (g) \ch{WSe2} monolayer from the \ch{Fe/WSe2/Pt} heterostructure; (h) Fe layer, (i) Pt layer, and (j) \ch{W_{0.9}V_{0.1}Se2}  from the \ch{Fe/W_{0.9}V_{0.1}Se2/Pt} HST. The band structures are given in the native Brillouin zones for each material. The number of Fermi-surface states is given by the color bar.}
    \label{fig:4}
    \end{center}
\end{figure}

To get a better understanding of the spin-resolved distribution across the interface, we also provide results for the computed Fermi Surfaces (FSs) for each HST component. The zone folding due to the reduced and incommensurate Brillouin zones in the supercell also appears in the FS for each HST. In order to monitor the evolution of the different carriers, the disentangled FSs for the majority and minority spins for each layer in Fe/Pt, \ch{Fe/WSe2/Pt} and \ch{Fe/W_{0.9}V_{0.1}Se2/Pt} HSTs in the $k_x-k_y$ plane are calculated. Fig. \ref{fig:4}a shows rather separated regions in Fe designated to the majority and minority spins, while for Pt in Fig. \ref{fig:4}b such a distinction is not much pronounced. Inserting the \ch{WSe2} layer changes the spin selection regions in $\mathbf{k}_{\perp}$ space for the Pt and Fe layers. Fig. \ref{fig:4}f,i show that there are no minority-spin states close to the $X (0.5\; 0 \; 0)$, $Y (0 \; 0.5 \; 0)$ point whereas majority-spin states exist almost everywhere in the surface Brillouin zone. 

We next examine the FS of the TMDC monolayer. Fig. \ref{fig:4}c, d shows that extended regions (around $K (1/3 \; 1/3 \; 0)$ points) are designated for both majority carriers despite the fact, that an isolated \ch{WSe2} does not have an inherent magnetism. The isolated V-doped monolayer, on the other hand, displays regions centered around the $K$-point for majority spins only, whose spread increases as the V-doping is increased (Fig. \ref{fig:4}c, d). The FS for the \ch{W_{0.9}V_{0.1}Se2} from the \ch{Fe/W_{0.9}V_{0.1}Se2/Pt} becomes very similar to the one for \ch{WSe2} from \ch{Fe/WSe2/Pt}, which shows that the spin transfer mechanism through the TMDC does not change significantly upon doping.

\begin{figure}[H]
    \begin{center}
    \includegraphics[width = 1 \textwidth]{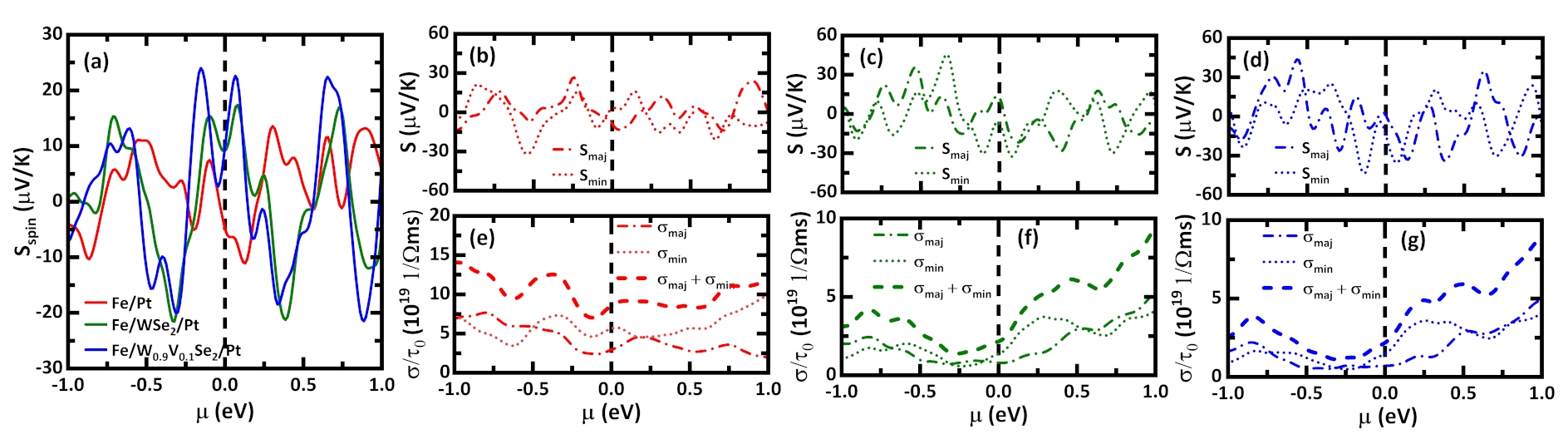}
    \caption{(a) The spin Seebeck coefficient $S_{spin}$ as a function of chemical potential $\mu$ for the Fe/Pt, \ch{Fe/WSe_2/Pt}, \ch{Fe/W_{0.9}V_{0.1}Se2/Pt} HSTs;   The majority $S_{maj}$ and minority $S_{min}$ Seebeck coefficients as a function of $\mu$ for (b) Fe/Pt, (c) \ch{Fe/WSe2/Pt}, (d) \ch{Fe/W_{0.9}V_{0.1}Se2/Pt} HSTs; The majority $\sigma_{maj}$, minority 
    $\sigma_{min}$ and total $\sigma_{maj}+\sigma_{min}$ electric conductivities divided by the constant relaxation time $\tau_0$ as a function of 
    $\mu$ for (b) Fe/Pt, (c) \ch{Fe/WSe2/Pt}, (d) \ch{Fe/W_{0.9}V_{0.1}Se2/Pt} HSTs. All results are at $T=300$ K.}
    \label{fig:5}
    \end{center}
\end{figure}

The spin dependent transport properties of the HSTs are directly related to their electronic structure. The spin Seebeck effect associated with the thermal injection of spin currents is of particular interest for spintronics and spin caloritronic devices, which rely on large spin Seebeck coefficients\cite{Adachi2013}.  Using linear response theory within the transport Boltzmann equation, we calculate the spin Seebeck coefficient 
$S_{spin}=\frac{\sigma_{maj}S_{maj}-\sigma_{min}S_{min}}{\sigma_{maj}+\sigma_{min}}$ 
($\sigma_{maj,min}, S_{maj,min}$ - electric conductivities and Seebeck coefficients for the majority and minority carriers) for several HSTs. Fig. \ref{fig:5}a shows that the magnitude of $S_{spin}$ at $T=300$ K for the HSTs with \ch{WSe2} is larger than the one for \ch{Fe/Pt} for practically all values of the chemical potential. For example, at $\mu=0$ $|S_{spin}|$ for \ch{Fe/W_{0.9}V_{0.1}Se2/Pt} is more than doubled the one for \ch{Fe/Pt}. In the peak-like features in the spin Seebeck coefficients at $\mu\sim -0.1$ eV and $\mu\sim -0.8$ eV, this increase is even strong as $|S_{spin}|$ for \ch{Fe/W_{0.9}V_{0.1}Se2/Pt} shows triple values compared to \ch{Fe/Pt}. 

One of the factors contributing to the overall enhancement of the spin Seebeck coefficient magnitude is the decreased $\sigma_{maj}+\sigma_{min}$ for the HSTs with the TMDC monolayer as seen in Fig. \ref{fig:5}e-g. The insertion of \ch{WSe2} leads to reduced charge transfer due to the larger separation between the Fe and Pt layers (Table \ref{tab:1}), which explains the decrease in the overall electronic transport for \ch{Fe/WSe2/Pt} and \ch{Fe/W_{0.9}V_{0.1}Se2/Pt}. Fig. \ref{fig:5}b-d further shows that there is larger disparity between $S_{maj}$ and $S_{min}$ for the HSTs with the TMDC as compared with \ch{Fe/Pt} for which the spin resolved Seebeck coefficients have similar magnitudes for most $\mu$. Our calculations also show that the enhanced difference between $|S_{maj}|$ and $|S_{min}|$ is attributed to the larger spin majority channels for HSTs with \ch{WSe2} as found from the FS distributions in Fig. \ref{fig:4}. Given that Seebeck coefficients are associated with localized features in DOS \cite{Popescu2010,Sofo1996}, the larger peaks in the majority DOS (Fig. \ref{fig:2}) is another factor contributing to the enhanced difference in the spin resolved Seebeck coefficients for the HSTs with TMDC monolayers.

Our first principles simulations show that  \ch{Fe/WSe2/Pt} HSTs exhibit complex electronic and magnetic properties. The pristine \ch{WSe2}, which is non magnetic, acquires magnetic moments on its atoms due to its proximity to the FM substrate of Fe. This distribution can be locally disturbed when defects or dopants are present. The chemically inert \ch{WSe2} weakens the bond between Fe and Pt, but significant charge transfer occurs through the monolayer in the HST. This rather stronger than vdW coupling may be seen in the DOS and band structure results showing the detailed orbital hybridization balance. The insertion of the TMDC monolayer also leads to an increased DOS at the Fermi level, which is about   $(23\%)$ higher than DOS of the reference HST. This is interesting to compare with  \ch{YIG/C60/Pt}: a related HST in which the \ch{C60} layer is also a semiconductor which bonds relatively strongly to YIG and Pt \cite{Kalappattil2020}. In this case, the six-fold DOS increase at the Fermi level is attributed to a much reduced conductivity mismatch, which explained the giant longitudinal spin Seebeck in \ch{YIG/C60/Pt} when compared with \ch{YIG/Pt} \cite{Kalappattil2020}. DOS increase in \ch{Fe/Pt} with TMDC monolayers is also present, which can be a factor responsible for the longitudinal spin Seebeck enhancement as reported in \cite{Dastgeer2019, Lee2020, Lee2020_a, Lee2021}. 

The unfolded band structures for the \ch{Fe/WSe2/Pt} HSTs are especially useful in understanding to what extent each component preserves its electronic properties. Of special interest is the \ch{WSe2} and what happens to its semiconducting nature after its insertion between the Fe and Pt substrates. Our calculations show that much of the ``identity'' of the TMDC monolayer is lost as there are numerous bands with various intensities crossing the Fermi level. The strong hybridization with orbitals from the substrates alters the character of the energy bands of the  \ch{WSe2}, which shows that its semiconducting behavior has changed. The presence of defects or dopants do not alter significantly this type of overall behavior. In light of this situation, potential experiments accessing the energy gap of the TMDC from the HST via optical means may need to be reevaluated \cite{Phan2021}.

The spin resolved atomic distribution and the FS calculations are also instrumental in understanding how localized spin is transferred across the interface of \ch{Fe/WSe2/Pt} HSTs. Due to the proximity of the Fe substrate,  \ch{WSe2} becomes ferromagnetically polarized. The Pt substrate also has magnetization $m_3\neq 0$. Furthermore, the overall magnetization of \ch{Fe/WSe2/Pt} is enhanced when comparing with the reference \ch{Fe/Pt} HST. The spin resolved FSs show large overlapping regions in $\mathbf{k_i}$ space for both spin carriers, and for the \ch{WSe2} there is little difference between the areas corresponding to the majority or minority carriers. As a result, we see that there is an increased density of magnetic atoms giving rise to a higher $M$, which effectively enhances the spin mixing and overall spin transfer in the HST as shown theoretically in \cite{Jia2011}. 

Our first principles simulations of \ch{Fe/WSe2/Pt} with pristine, defective or V-doped TMDC monolayer reveal complex electronic and magnetic properties of the HSTs and their individual components. Although the ferromagnetism in \ch{WSe2} can be tuned via doping and/or defects, we find that this effect becomes secondary in the overall behavior of the HST. The semiconducting nature of the TMDC becomes diluted when the monolayer is between the Fe and Pt substrates. The total DOS increase at the Fermi level, larger magnetization, and importantly much enhanced spin density at the interface with overlapping majority and minority regions in the FS of \ch{Fe/WSe2/Pt} HST influence the electronic and spin transport across the interface. The combined effect of all of these factors results in larger spin Seebeck coefficients in HSTs with TMDC layers, which can be useful in interpreting recent experiments reporting enhanced magnetization and spin resolved transport properties \cite{Yun2020, Pham2020, Zhang2020, Dastgeer2019, Lee2020, Lee2020_a, Lee2021, Wang2015, Nguyen2021, Yun2022}. 

{\it Computational Methods:} The first principles simulations are performed with the DFT Vienna ab initio Simulations Package (VASP), which is based on projector-augmented wave (PAW) pseudopotentials and periodic boundary conditions \cite{Kresse1996, Kresse1996_b}. The exchange-correlation effects are considered using the Perdew-Burke-Ernzerhof (PBE) generalized gradient approximation with an included spin-orbit coupling (SOC) \cite{Perdew1996}. The vdW interaction is included via the vdW-D3 (BJ) dispersion correction \cite{Grimme2011}. The electronic shell structure is taken to have 8 valence electrons of Fe (3\ch{d^6} 4\ch{s^2}), 6 valence electrons of W (6\ch{s^2} 5\ch{d^4}), 6 valence electrons of Se (4\ch{s^2} 4\ch{p^4}) and 10 valence electrons of Pt (5\ch{d^9} 6\ch{s^1}). The kinetic energy cutoff is $450$ eV. In the construction of the HST supercells, we use the rectangular unit cell of \ch{WSe_2} $(a = 3.32 \text{ \AA}, b = 5.75 \text{ \AA})$ with a $6 \times 6$ Fe supercell (108 Fe atoms), a $3 \sqrt{3} \times 5$ \ch{WSe_2} supercell ($30$ W and $60$ Se atoms), a $3 \sqrt{2} \times 3 \sqrt{2}$ Pt supercell ($108$ Pt atoms). To eliminate the interactions between periodic images, a vacuum layer of $15$ \text{\AA} is taken between the HST layers for all calculations. 
The relaxation criteria of all systems is performed by allowing all cell parameters to change with $10^{-5}$  eV energy and $0.01$ eV⁄\text{\AA} force relaxation criteria. The intergration over the Brillouin zone is performed at a single $\Gamma$ point for the optimization calculation. The DOS, charge density differences and Bader charges calculations are accomplished with $2 \times 2 \times 1$ $\Gamma$-centered Monkhorst-Pack k-point mesh. These calculations are performed with employing the Gaussian smearing with 0.1 eV width. To account for the strongly correlated electronic systems, we employed the $GGA+U$ method within Dudarev’s approach with an on-site Coulomb correction in the localized d orbitals of the V atoms. The parameter value of $U_{eff} = U - J = 3.0$ eV is selected, which has been shown to describe the ground-state electronic and magnetic properties of the system accurately \cite{Duong2019, Yun2020}. For the Fe, W, Se and Pt, $U_{eff} = 0$ eV. The visualization of all systems is attained in VESTA \cite{Momma2011}.

Given the zone folding problem that occurs in simulations involving supercells, the disentanglement of the various band structures is performed by adapting the vaspkit code \cite{Wang2021}. Within this code, one creates $\mathbf{k_i}$ vectors along the high symmetry directions in the primitive Brillouin zone and then translates them into the supercell reciprocal space. Performing VASP calculation using these $\mathbf{k_i}$ vectors yields the plane wave coefficients and eigenvalues of each state necessary to determine spectral weights for the projection of the supercell eigenstates on the primitive cell eigenstates.

The FS unfolding is achieved using the bands4vasp package \cite{Dirnberger2020}, in which  the energy eigenvalues for every $\mathbf{k_i}$ primitive cell vector are assigned to a particular energy band by considering similarities in the orbital symmetry of the eigenstates. Then, a band center of mass is defined by averaging the energy values with weights given by the corresponding $P_{ \mathbf{K} m} ( \mathbf{k_i})$ character corresponding to the  projection of the supercell eigenstates $\left| \mathbf{K} m \right>$ on the primitive cell eigenstate $\left| \mathbf{k} n \right>$. Finally, the intersection of every band with the Fermi level is found by interpolation. Because the band structure and FS unfolding are computationally expensive, we consider the supercells consisting of $2 \times 6$ Fe unit cells (36 atoms), $\sqrt{3} \times 5$ unit cells for the  monolayer (30 atoms), and $\sqrt{2} \times 3\sqrt{2}$ Pt unit cells (36 atoms) are constructed. The optimization calculations of these system are performed with $6 \times 2 \times 1$ $\Gamma$-centered Monkhorst-Pack k-point mesh. 

The transport properties of the HSTs are calculated with the BoltzTraP2 code \cite{Madsen2018} with a large k-mesh of 112 000 points for more accurate results. The code utilizes linear response theory of the Boltzmann equation within a constant relaxation time approximation. In general, this is a good approximation as long as the energy changes are at a smaller scale as compared to the energy changes in the density of states, which is typically the case for metals and many semiconductors.

\begin{acknowledgement}
D. T.-X. D. would like to acknowledge support from the Presidential Fellowship sponsored by the University of South Florida. L.M.W. acknowledges financial support from the US Department of Energy under Grant No.DE-FG02-06ER46297. Computational resources are provided by USF Research Computing.
\end{acknowledgement}

\begin{suppinfo}
The Supporting Information contains structural data for the strains in the considered heterostructures upon the constructuion their supercells. Results for the calculated unfolded band structures of Fe/Pt, \ch{Fe/WSe2}, \ch{Pt/WSe2}, \ch{Fe/WSe2/Pt} and \ch{Fe/W_{0.9}V_{0.1}Se2/Pt} heterostructures and their constituents are also shown for further comparison with the results presented in the main text.

The Supporting Information is available free of charge at \url{https://pubs.acs.org/doi/10.1021/acs.jpclett.2c01925}.
\end{suppinfo}

\bibliography{ref}

\end{document}